\documentclass[seceq]{ptptex}

\newcommand{\GB}{\mathcal{G}}
\newcommand{\de}{\mathrm{d}}

\title{
Linear growth of matter density perturbations in $f(R,\GB)$ theories
}

\author{
Antonio \textsc{De Felice}$^{1,}$ 
and Teruaki \textsc{Suyama}$^{2,}$
}

\inst{
$^1$Department of Physics, Faculty of Science, Tokyo University of Science, 1-3, Kagurazara,
Shinjuku-ku, Tokyo 162-8601, Japan\\
$^2$Research Center for the Early Universe, The University of Tokyo, 4-3-1, Hongo, Bunkyo-ku, Tokyo 113-0033, Japan
}

\abst{
We derive the equation of matter density perturbations on sub-horizon scales around 
a flat Friedmann-Lema\^\i tre-Robertson-Walker background for the general
Lagrangian density $f(R,\GB)$ that is a function of a Ricci scalar $R$ and a Gauss-Bonnet term $\GB$.
We find that the effective gravitational constant generically scales as distance squared at small distances.
The effect of this diminishing of the gravitational constant might be important in the gravitational dynamics 
of cosmic objects such as galaxies, which can be in principle tested by observations.
We also provide the general expressions for the effective anisotropic stress, 
which is useful to constrain modified gravity models from observations of large-scale structure and weak lensing.
We also find that there is a special class of theories which 
evade this unusual behaviour and that the condition 
to belong to this special class is exactly the same as the one 
for not having super-luminal modes with propagation speed proportional to their wavenumber.
}

\begin{document}

\maketitle

\section{Introduction}

Modifying the law of gravity is a well-known possible explanation for
the origin of the accelerated expansion of the universe (for a general
review of the acceleration mechanism, see
e.g.~\cite{Copeland:2006wr,DeFelice:2010aj}).  Precise observations of
cosmic structures are now becoming powerful methods to distinguish
modified gravity scenarios from other scenarios such as a cosmological
constant.  The basic observation which enables us to hope it is indeed
feasible to recognize among these theories the right one, is that
modified gravity theories generically predict characteristic
scale-dependent growth of the matter density perturbations, which thus
leaves distinct patterns of the cosmic structures. In other words, the
dynamics of matter perturbation becomes scale dependent, and each
scale feels in general a different effective Newtonian constant, i.e.\
a different gravity law. Therefore by precisely measuring the patterns
of the cosmic structures, we can in principle accept or rule out the
modified gravity scenarios as the origin of the accelerated expansion.
There have been a number of recent studies about the evolution of
density perturbations for various types of the modified gravity
theories, e.g.~for $f(R)$ and scalar-tensor theories
\cite{Boisseau:2000pr,EspositoFarese:2000ij,Zhang:2005vt,Carroll:2006jn,Song:2006ej,Bean:2006up,Li:2007xn,Sawicki:2007tf,Uddin:2007gj,Tsujikawa:2007gd,Pogosian:2007sw,Gannouji:2008wt,Nesseris:2008mq,Motohashi:2009qn,Narikawa:2009ux}
for $R+f(\GB)$ theories \cite{DeFelice:2009rw,Li:2007jm}, for
$f(R,R_{\mu \nu}R^{\mu \nu},\Box R)$ theories \cite{Nesseris:2009jf}, and for the DGP-inspired Galileon field \cite{DeFelice:2010pv}.

In \cite{DeFelice:2010sh}, we have studied the basic properties of
propagation of the perturbations around a flat Friedmann-Lema\^\i
tre-Robertson-Walker (FLRW) background for the general $f(R,\GB)$
theories with a perfect fluid.  There, we found that there are four
independent modes for the perturbations. Two of which have dispersion
relations like $\omega^2 =A \Xi k^4+{\cal O}(k^2)$, where $A$ is a background dependent 
quantity and $\Xi$ is a determinant of the Hessian of $f(R,\GB)$:
\begin{equation}
\Xi \equiv \frac{\partial^2 f}{\partial R^2}\frac{\partial^2 f}{\partial \GB^2} -{\left( \frac{\partial^2f}{\partial R \partial \GB} \right)}^2,
\end{equation}
evaluated for the background metric.
For the special theories such as $f(R)$ and $R+f(\GB)$ that give vanishing $\Xi$,
the leading term in the dispersion relation starts from $k^2$ order.
On the other hand, the remaining two have
the standard ones $\omega^2 =c_s^2 k^2$, where $c_s^2 ={\dot P}/{\dot
  \rho}$ is the sound speed of the fluid.  The former generally
results in super-luminal propagation for short wavelength modes, while
the latter is the standard dispersion relation for a perfect fluid.

Based on the basic studies in \cite{DeFelice:2010sh}, in this paper we
restrict ourselves to the non-relativistic matter ($c_s=0$) and
study in detail the properties of the matter density perturbation for
$f(R,\GB)$ theories to derive the consequences that are linked with
observations.  We will provide general formulae for the parameters
that are now widely used in literature to parametrize the deviation
from GR for $f(R,\GB)$ theories.  Interestingly, it turns out that the
effective gravitational constant generically decays at short distance
$r$, obeying a scaling $\propto r^2$.  We also discuss a class of
theories which evade this diminishing of the gravitational constant.

\section{Modified Gravity Models}
We consider the $f(R,\GB)$ theories where cold dark matter(CDM) and
baryons are minimally coupled to gravity. The corresponding action is
given by
\begin{eqnarray}
S=\frac{1}{16\pi G_N} \int d^4x\sqrt{-g}~f(R,\GB)+\int d^4x \sqrt{-g}~ {\cal L}_m, \label{action0}
\end{eqnarray}
where $G_N$ is the Newton's constant, $R$ is the Ricci scalar and
$\GB$ is the so-called Gauss-Bonnet term defined by
\begin{eqnarray}
\GB\equiv R^2-4R_{\mu \nu} R^{\mu \nu}+R_{\mu \nu \alpha \beta} R^{\mu \nu \alpha \beta},
\end{eqnarray}
and ${\cal L}_m$ is the Lagrangian density for the CDM and baryons.

The equations of motion for $g_{\mu \nu}$ are given by
\begin{eqnarray}
R_{\mu \nu}-\tfrac{1}{2} g_{\mu \nu}R-\Sigma_{\mu \nu}=
\frac{8\pi G_N}{F}\, T_{\mu\nu}. \label{eom}
\end{eqnarray}
Here $\Sigma_{\mu \nu}$ is the effective energy momentum tensor
defined by
\begin{eqnarray} 
\Sigma_{\mu \nu}&=&\frac1F \bigl(\nabla_\mu \nabla_\nu F-g_{\mu \nu} \Box F+2R \nabla_\mu \nabla_\nu \xi-2g_{\mu \nu} R \Box \xi-4R_\mu^{~\lambda} \nabla_\lambda \nabla_\nu \xi \nonumber -4R_\nu^{~\lambda} \nabla_\lambda \nabla_\mu \xi \\ 
&&+4R_{\mu \nu} \Box \xi+4 g_{\mu \nu} R^{\alpha \beta} \nabla_\alpha \nabla_\beta \xi+4R_{\mu \alpha \beta \nu} \nabla^\alpha \nabla^\beta \xi-\tfrac{1}{2}\,g_{\mu \nu} V \bigr), \label{effective-energy-momentum}
\label{enemom}
\end{eqnarray}
where $F$, $\xi$ and $V$ are defined by
\begin{align}
&F \equiv \frac{\partial f}{\partial R}, ~~~~~\xi \equiv \frac{\partial f}{\partial \GB}, \\
&V \equiv RF+\GB \xi-f.
\end{align}
$T_{\mu \nu}$ is the energy-momentum tensor for CDM and baryons.
It is given by 
\begin{equation}
T_{\mu \nu}=\rho^{(b)} u_\mu^{(b)} u_\nu^{(b)}+\rho^{(c)} u_\mu^{(c)} u_\nu^{(c)},
\end{equation}
where $b/c$ represents baryons/CDM respectively.
As usual, we impose a normalization condition $u_\mu^{(A)} u^{\mu (A)}=-1$ for each component.

\subsection{Background dynamics}
We assume that the background spacetime is a flat Friedmann-Lema\^\i tre-Robertson-Walker (FLRW) universe whose metric is given by
\begin{eqnarray}
\de s^2=-\de t^2+a^2(t)\, \delta_{ij}\de x^i\, \de x^j. \label{bg1}
\end{eqnarray}
Then we find that the background equations are given by
\begin{align}
3H^2&=\frac{1}{F} \left[\tfrac12\, V-3H{\dot F}-12H^3 {\dot \xi} \right]+\frac{8\pi G_N \rho_m}{F},\\
{\ddot F}&=-8\pi G_N \rho_m+H {\dot F}-2 {\dot H}F+4H^3 {\dot \xi}-8 H {\dot H} {\dot \xi}-4H^2 {\ddot \xi},
\end{align}
where $\rho_m = \rho^{(b)}+\rho^{(c)}$ is the total matter density.

\subsection{Scalar-type Perturbation}

We consider linear scalar-type perturbations around the metric Eq.~(\ref{bg1}).
We always work in the Newtonian gauge:
\begin{eqnarray}
\de s^2=-(1+2\Psi)\, \de t^2+a^2(t) (1+2\Phi) \delta_{ij}\, \de x^i \,\de x^j.\label{per1}
\end{eqnarray}
We also decompose the energy-momentum tensor for the CDM and baryons as
\begin{equation}
\rho^{(A)} \to \rho^{(A)}+ \delta \rho^{(A)}=\rho^{(A)} (1+\delta^{(A)}), ~~~~~u_i^{(A)} \to \partial_i V^{(A)},~~~~~(A=b,c).
\end{equation}
All perturbation variables for the matters are also defined in the Newtonian gauge.

Then the perturbation equations, in Fourier space, are given by
\begin{align}
&3H^2 \Psi-\frac{k^2}{a^2} \Phi-3H{\dot \Phi}=\frac{1}{2F} \bigg[ 3({\dot F}+12H^2 {\dot \xi}) {\dot \Phi}+8 \frac{k^2}{a^2} H {\dot \xi} \Phi-6H ({\dot F}+8 H^2 {\dot \xi}) \Psi \nonumber \\
&\hspace{40mm}+3H ({\dot {\delta F}}+4H^2 {\dot {\delta \xi}})+\left( -3(H^2+{\dot H})+\frac{k^2}{a^2} \right) (\delta F+4H^2\delta \xi) \bigg] \nonumber \\
&\hspace{40mm}-\frac{4\pi G_N( \delta \rho^{(b)} +\delta \rho^{(c)}) }{F}, \label{per00} \\
&{\dot \Phi}-H\Psi =\frac{1}{2F} \bigg[H\delta F-{\dot {\delta F}}-4H^2 {\dot {\delta \xi}}+4H^3 \delta \xi+({\dot F}+12H^2 {\dot \xi}) \Psi-8H {\dot \xi} {\dot \Phi} \bigg] \nonumber \\
&\hspace{20mm}+\frac{4\pi G_N(\rho^{(b)} V^{(b)}+\rho^{(c)} V^{(c)})}{F}, \label{per0i}\\
&\Phi+\Psi=-\frac{1}{F} \bigg[ 4H{\dot \xi}\Psi+4{\ddot \xi} \Phi+\delta F+4(H^2+{\dot H}) \delta \xi \bigg],\label{pertrls}\\
&{\dot \delta^{(b)}}+3{\dot \Phi}-\frac{k^2}{a^2}V^{(b)}=0, \label{0baryon} \\
&{\dot \delta^{(c)}}+3{\dot \Phi}-\frac{k^2}{a^2}V^{(c)}=0, \label{0cdm} \\
&{\dot V}^{(b)}+\Psi=0, \label{ibaryon} \\
&{\dot V}^{(c)}+\Psi=0. \label{icdm}
\end{align}
The perturbation quantities $\delta F$ and $\delta \xi$ are related to
perturbations of Ricci scalar and the Gauss-Bonnet term through the
equations:
\begin{align}
\delta F=&F_R \delta R +F_{\GB} \delta \GB, \label{Fcon}\\
\delta \xi=&F_{\GB} \delta R+\xi_{\GB} \delta \GB, \label{Gcon}
\end{align}
where $F_R = \partial F / \partial R$ and so on.  In the second
equation, we have used an identity $\xi_R=F_{\GB}$.  
By perturbing the scalars $R$ and $\GB$, we have the following relations:
\begin{align}
\delta R&=-2 \left[ 6(2H^2+{\dot H}) \Psi-2\frac{k^2}{a^2} \Phi-\frac{k^2}{a^2} \Psi-3{\ddot \Phi}-12H{\dot \Phi}+3H{\dot \Psi} \right], \label{perricci}\\
\delta \GB&=-8 \bigg[ 12(H^2+{\dot H}) H^2 \Psi-3H^2 {\ddot \Phi}+3H^3 {\dot \Psi} -6H(2H^2+{\dot H}) {\dot \Phi} \nonumber \\
&\hspace{10mm}-2\frac{k^2}{a^2} (H^2+{\dot H}) \Phi-\frac{k^2}{a^2} H^2 \Psi \bigg]. \label{pergb}
\end{align}

\section{Properties of linear growth in $f(R,\GB)$ theories}
\subsection{Catch-up of baryons}

Because the matters are minimally coupled to gravity, the baryon
density perturbation starts catching up CDM after the
recombination. To explain this behavior in more detail, let us
introduce a quantity $\Delta =\delta^{(b)}-\delta^{(c)}$.  By using
(\ref{0baryon})-(\ref{icdm}), we can derive a following evolution
equation for $\Delta$,
\begin{equation}
{\ddot \Delta} +2H {\dot \Delta}=0.
\end{equation}
The general solution of this differential equation is given by
\begin{equation}
\Delta (t)=a^2(t_1) {\dot \Delta} (t_1) \int_{t_1}^t \frac{dt'}{a^2(t')} +\Delta (t_1),
\label{catch-sol}
\end{equation}
where $t_1$ is an arbitrary constant and the two initial conditions
$\Delta(t_1)$ and $\dot\Delta(t_1)$ may be in general $k$
dependent. If we take $t_1$ as a time just after the decoupling of CMB
photons from baryons, the expansion of the universe can be well
approximated by the one for the matter dominated universe, i.e., $a(t)
\propto t^{2/3}$.
Then the integral on the RHS of Eq.\
(\ref{catch-sol}) is mostly determined by its lower limit which
becomes
\begin{equation}
a^2(t_1) {\dot \Delta} (t_1) \int_{t_1}^t \frac{dt'}{a^2(t')} = {\cal O} \left( \frac{{\dot \Delta} (t_1)}{ H(t_1)} \right).
\end{equation}
Since both matter and baryon density perturbations can grow only on
the cosmological time scale $\sim 1/H(t_1)$, the first term in
(\ref{catch-sol}) remains of the same order of magnitude as the second
one even at late times.  Meanwhile the matter density perturbations
themselves grow from tiny fluctuations to more than ${\cal O}(1)$ to
make cosmic structures.  Therefore, a relative difference $\Delta /
\delta^{(c)}$ quickly decays and the baryon density perturbation
becomes almost the same as the CDM density perturbation. Note that we
have used only the conservation equations for the matter fields to derive this
result.  The only place where the effect of modification of gravity
enters is the evolution of the scale factor appearing in the integral
in (\ref{catch-sol}) due to the modification of the Friedmann
equation.  However, since the effects of modification of gravity are
negligible at high cosmological redshift, we expect that the evolution
of $\Delta$ to change only slightly from the one in GR.

At late times, when the modification of gravity becomes important, we have $\delta^{(b)} \approx \delta^{(c)}$.
Then it is a good approximation to consider a total density perturbation $\delta_m$ defined by
\begin{equation}
\delta_m \equiv \frac{\rho^{(b)} \delta^{(b)}+\rho^{(c)} \delta^{(c)}}{\rho_m} \approx \delta^{(b)} \approx \delta^{(c)},
\end{equation}
instead of considering $\delta^{(b)}$ and $\delta^{(c)}$
independently.  The approximation $\delta_m=\delta^{(b)}=\delta^{(c)}$
yields another relation $V^{(b)}=V^{(c)} \equiv V_m$, which can be
derived by substracting (\ref{0baryon}) from (\ref{0cdm}).

To conclude, baryon density perturbation catches up the CDM density
perturbation in the same way as in GR.  Correspondingly, we can safely
replace the conservation equations (\ref{0baryon})-(\ref{icdm}) by
\begin{align} 
&{\dot \delta_m}+3{\dot \Phi}-\frac{k^2}{a^2}V_m=0, \label{0m} \\
&{\dot V_m}+\Psi=0, \label{im} 
\end{align}
to study structure formation at low redshift universe.
In the following analysis, we use $\delta_m$ and $V_m$ for the matter density perturbation.

\subsection{Effective gravitational constant}

One generic feature of the modified gravity models is that the effective
gravitational constant has a scale dependence over cosmological scales.
In the literature it is common to define the effective gravitational constant by a Poisson equation \footnote{
Some literature (e.g. \cite{Song:2010rm,Daniel:2010ky}) define the effective gravitational constant by using $\Phi$ instead
of $\Psi$. If we use $\Phi$, the effective gravitational constant then represents how strongly the
space is curved by the presence of matter. To go from one to the another, we have to multiply or divide by 
a factor $\Phi/\Psi$.}:
\begin{equation}
\frac{k^2}{a^2} \Psi = -4 \pi G_N Q(k,t) \rho_m \delta_m, \label{Poisson}
\end{equation}
where we parametrize the deviation from GR by $Q(k,t)$ (as in GR we
have $Q=1$).  The modification of the Poisson equation will affect the
evolution of the matter density perturbations, which can be tested by
cosmological observations such as weak lensing.  We use $\Psi$, the
time component of the metric perturbation, to define the effective
gravitational constant.  Therefore the gravitational constant obtained
by measuring a force between two mass points is given by $G_NQ$.

In this subsection, we derive the general expression of $Q(k,t)$ for
$f(R,\GB)$ theories on sub-horizon scales which are relevant to
cosmological observations.  To this end, following the approach in
\cite{Boisseau:2000pr,EspositoFarese:2000ij,Tsujikawa:2007gd,Nesseris:2009jf},
we will use a subhorizon approximation under which we assume that time
derivative of any perturbation variable is much less than its spatial
derivative or equivalently, in terms of Fourier space, a term
multiplied by $k$.  Therefore if there are time derivatives of a
perturbation variable in addition to terms of the same variable
multiplied by $k^2$ in a perturbation equation, we drop the first from
the equation.  By this procedure, we are dropping the two fast
oscillating modes among the four whose dispersion relation is given by
$\omega^2 = A\Xi k^4$ ($A \Xi$ must be
positive to avoid an instability of the FLRW universe) and are taking
into account only the remaining two modes which evolve on cosmological
time scale.

With the sub-horizon approximation, the perturbation equations are given by
\begin{align}
&-\frac{k^2}{a^2} \Phi=\frac{k^2}{2Fa^2} \left( 8  H {\dot \xi} \Phi+ \delta F+4H^2\delta \xi \right)-\frac{4\pi G_N \rho_m \delta_m }{F}, \label{eff00} \\
&\Phi+\Psi=-\frac{1}{F} \bigg[ 4H{\dot \xi}\Psi+4{\ddot \xi} \Phi+\delta F+4(H^2+{\dot H}) \delta \xi \bigg], \label{efftrls}\\
&\delta R=2 \frac{k^2}{a^2} \left( 2 \Phi+ \Psi \right), \label{effricci}\\
&\delta \GB=8\frac{k^2}{a^2} \bigg[ 2 (H^2+{\dot H}) \Phi+ H^2 \Psi \bigg], \label{effgb} \\
&{\dot \delta_m}-\frac{k^2}{a^2}V_m=0, \label{eff0} \\
&{\dot V_m}+\Psi=0, \label{effi}
\end{align}
where we again use the relations (\ref{Fcon}) and (\ref{Gcon}) to
relate $(\delta F, \delta \xi)$ with $(\delta R, \delta \GB)$. In Eq.\ (\ref{eff00}) we have neglected the contributions of the $\Psi$ field as it is of the same order of $\Phi$. This assumption turns out to be consistent when we write, as we shall see later on, $\Phi$ in terms of $\Psi$.  We do
not include (\ref{per0i}) as it is not necessary for our present
purpose.

To derive the effective gravitational constant, we first combine
(\ref{Fcon}), (\ref{Gcon}), (\ref{effricci}) and (\ref{effgb}) to
express $\delta F$ and $\delta \xi$ in terms of $\Phi$ and $\Psi$.
Substituting these results into (\ref{efftrls}), we obtain an equation
which only contains $\Phi$ and $\Psi$.  By solving the obtained
equation with respect to $\Phi$, $\Phi$ is expressed solely by $\Psi$,
and at this point we can check the consistency of the assumption we
considered earlier. By putting this relation back into equations for
$\delta F$ and $\delta \xi$ to eliminate $\Phi$, both $\delta F$ and
$\delta \xi$ are also expressed solely by $\Psi$.  Then by combining
these results with (\ref{eff00}), we find that $\delta_m$ can be also
written solely by $\Psi$, from which we can derive the desired
expression for $Q(k,t)$:
\begin{equation}
Q(k,t)=\frac{A_1+A_2 {[k/(aH)]}^2}{B_1+B_2{[k/(aH)]}^2+B_3 {[k/(aH)]}^4}, \label{eqQ}
\end{equation}
where time-dependent coefficients are given by
\begin{align}
&A_1=F+4\ddot\xi, \\
&A_2=4H^2 \bigg[  F_R+8F_{\GB} (H^2+{\dot H})+16 \xi_{\GB} {(H^2+{\dot H})}^2 \bigg], \\
&B_1=(F+4H\dot\xi)^2, \\
&B_2=H^2( 3F+16H {\dot \xi}-4{\ddot \xi} ) [ F_R+4F_\GB (2H^2+{\dot H})+16H^2 \xi_\GB (H^2+{\dot H})] \nonumber \\
&\hspace{10mm}+4 {\dot H} H^2 (F+4{\ddot \xi}) (F_{\GB}+4H^2 \xi_{\GB}), \\
&B_3=64 {\dot H}^2 H^4 (F_{\GB}^2-F_R \xi_{\GB})=-64 {\dot H}^2H^4 \Xi\,.
\end{align}
In $A_1$ and $B_1$, there are other terms that are functions of
$F_R,F_\GB$ and $\xi_\GB$.  However, they are subdominant compared to
$F$ terms and we have neglected them in the above equations.
We find that $B_3$ is proportional to $\Xi$. Therefore, it is absent for
the special models where $\Xi=0$.
As we will see, the behavior of the effective gravitational constant crucially
depends on if the $B_3$ is zero or not.

Let us consider $f(R)$ theories just to check that our result correctly reproduces $Q$ given
in the literature.
In this case, ${\dot \xi}={\ddot \xi}=F_{\GB}=\xi_{\GB}=0$ and we have 
\begin{equation}
Q=\frac{F+4F_RH^2 {[k/(aH)]}^2}{F^2+3F F_R H^2 {[k/(aH)]}^2}.
\end{equation}
This coincides with the one given in \cite{Tsujikawa:2007gd}.

\subsection{Interpretation of the result}
Now let us consider (\ref{eqQ}) where we find some new interesting consequences.  
For long wavelength modes $(k \to 0)$, we find that $Q$ becomes independent of $k$:
\begin{equation}
Q \approx \frac{A_1}{B_1}=\frac{c_{TT}^2}{2Q_{TT}}, \label{smallk:Q}
\end{equation}
where
\begin{equation}
  \label{eq:cTTQTT}
  c_{TT}^2=\frac{F+4\ddot\xi}{F+4H\dot\xi}\,,\qquad{\rm and}\qquad
  Q_{TT}=\tfrac12\,F+2H\dot\xi\, ,
\end{equation}
are respectively the squared speed and the coefficient of the kinetic
term for the gravitational wave modes of the theory (see
\cite{DeFelice:2009ak}). Both these quantities must be positive in
order to avoid ghosts and small scale Laplacian instabilities
($c_{TT}^2<0$). A consequence of Eq.\ (\ref{smallk:Q}) is that an
opposite sign for $Q_{TT}$ would lead to an effective repulsive
gravity law.
For short wavelength modes $(k \to \infty)$, we find that $Q$ is given by
\begin{equation}
Q \approx \frac{A_2}{B_3}\,\frac{a^2H^2}{k^2}.
\end{equation}
We see that $Q$ is inversely proportional to $k^2$, which means that
gravitational constant decays on small scales. This is quite
different from $f(R)$ theories because in $f(R)$, $Q$ approaches a
value $4/(3F)$ on small scales. Let us just refer \cite{Nesseris:2009jf}
which found that $k^{-2}$ scaling of the effective gravitational
constant is present for $f(R,R_{\mu \nu}R^{\mu \nu},\Box R)$ theories.
This theory does not overlap with $f(R,\GB)$ theories considered in
this paper as a quantity $R_{\mu \nu \alpha \beta} R^{\mu \nu \alpha
  \beta}$ does not enter in their action. Furthermore, that theory in general may contain spin-2 ghost degrees of freedom.

To see what is going on in real space, let us consider the gravitational potential
sourced by a local distribution of matter density on the cosmological background.
As the simplest case, let us consider a situation that a mass point is put at the origin,
i.e., $\delta \rho_m ({\vec x})=M \delta ({\vec x})$\ \footnote{This ansatz does not solve Eq.\ (\ref{evo}). However, we can give it as an initial condition for $\delta_m$ together with $\dot\delta_m=0$. The fact that for large $k$ $Q\propto k^{-2}$ implies that $\delta_m$ will not evolve rapidly but only in Hubble time. In other words, we can consider it as a good approximation in a short time range.}. Then the solution of the equation (\ref{Poisson}) in real space is given by
\begin{equation}
\Psi ({\vec x},t)=-\frac{2G_NM}{\pi r}\int_0^\infty \frac{d k}{k} \frac{A_1+A_2 {[k/(aH)]}^2}{B_1+B_2{[k/(aH)]}^2+B_3 {[k/(aH)]}^4} \sin (kr), \label{poten}
\end{equation}
where $r \equiv |{\vec x}|$.  Although we can do the integration
analytically, the result is rather complicated.  Here we decide to
provide only asymptotic behavior of (\ref{poten}) at large and small
distances from the source.

For large $r$, the dominant contribution to the integration comes from
a region near $k = 0$.  Then we can drop the higher order terms in $k$
appearing in the integrand and we have
\begin{equation}
\Psi ({\vec x},t)= -\frac{G_NM}{r}\,\frac{c_{TT}^2}{2Q_{TT}}+{\cal O} \left( \frac{1}{r^2} \right).
\label{eq:psir}
\end{equation}
As expected, the gravitational potential is inversely
proportional to $r$ with its amplitude multiplied by background
quantities. On the other hand, for short distance, the higher order
terms in $k$ become important and we have
\begin{equation}
\Psi ({\vec x},t) = {\rm const.}+\frac{G_NM A_2 a^2 H^2}{2B_3}r +{\cal O}(r^2).
\end{equation}
We find that both gravitational potential and gravitational force
approach constant values, which is consistent with the observation in
Fourier space that effective gravitational constant decays like
$\propto k^{-2}$ at large $k$.

It is important to notice that in the limit $A_1 \to 1$, $A_2 \to 0$,
$B_1 \to 1$, $B_2\to0$ and $B_3\to0$ one recovers the known GR result,
that is $Q=1$. Although we want to be as much general as possible in
this study, we notice that this happens for theories for which
$F_R,F_{\GB},\xi_{\GB}$ all go to 0 at early times. On the other hand,
all these terms are supposed to be of the same order at late times. At
the typical scales at which linear perturbation theory can be used to
test a theory, $k/(a_0 H_0)\sim100$, one can set some constraints as
follows
\begin{enumerate}
\item $|B_3/B_2|\times 10^4\lesssim 1$. This case happens when the
  modifications of gravity are either of the $f(R)$ kind or very weak
  even today, or when the expansion of the universe is almost de
  Sitter ($\dot H\approx0$). In particular the theory reduces to the
  special cases (for which $F_{\GB}^2-F_R \xi_{\GB}\approx0$)
  discussed later on. This case although possible is not interesting
  enough, because one would need some other experiment to disentangle
  these kind of theories from, say, the $f(R)$ ones.
\item $|B_3/B_2|\times 10^4\lesssim 1$ and $|B_2/B_1|\times
  10^4\lesssim 1$.  When this case happens, since $A_2$ is typically
  the same order of magnitude as $B_2$ unless the model parameters are
  fine-tuned, in general we also have $|A_2/A_1|\times 10^4\lesssim
  1$. This case leads to a matter spectrum close to GR.
\end{enumerate}
It is then evident that when any of the previous two conditions are
not satisfied we are in the full Modified Gravity regime, for which
bounds can be set. In particular for scales at which $B_3$ dominates,
since $Q\to0$ we would expect the matter power spectrum to reduce, as
gravity is weaker at those scales. On the other hand, at scales for
which $B_2$ is the dominant term, the quantity one needs to extract
from the particular model at hand is $Q\approx A_2/B_2$. If this
number is larger or smaller than $1/F$ then one obtains an enhancement
or a suppression of the matter power-spectrum. Finally, in general one
may find both these two different behaviors, which depend on the
scale.

\subsection{Evolution equation for the matter density perturbation}
From (\ref{eff0}) and (\ref{effi}), the following evolution equation
for $\delta_m$ is derived:
\begin{equation}
{\ddot \delta_m}+2H {\dot \delta_m}+\frac{k^2}{a^2} \Psi=0.
\end{equation}
Putting the Poisson equation (\ref{Poisson}) into this equation yields
a closed evolution equation for $\delta_m$,
\begin{equation}
{\ddot \delta_m}+2H {\dot \delta_m}-4\pi G_N Q \rho_m \delta_m =0. \label{evo}
\end{equation}

\subsection{Parameter $\Sigma$}
In this subsection, we provide expressions of a parameter $\Sigma$
that are sensitive to weak lensing for $f(R,\GB)$ theories.  This
parameter is defined by
\begin{equation}
\Sigma \equiv \frac{Q}{2} ( 1+\eta ),
\end{equation}
where $\eta$ is defined by
\begin{equation}
\eta \equiv -\frac{\Phi}{\Psi}.
\end{equation}
The photon propagation is sensitive only to a so-called lensing
potential $\Phi-\Psi$ and the parameter $\Sigma$ appears when we
rewrite $\Phi-\Psi$ in terms of $\delta_m$.  Therefore, $\Sigma$
parametrizes the relation between the lensing potential and the
density perturbation.  We find that $\Sigma$ is given by
\begin{equation}
\Sigma = \frac{C_1+C_2{(k/a)}^2}{B_1+B_2 {(k/a)}^2+B_3 {(k/a)}^4}, \label{sigma}
\end{equation}
where $C_1$ and $C_2$ are given by
\begin{align}
&C_1=F, \\
&C_2=3 \bigg[ F_R+4F_\GB (2H^2+{\dot H})+16H^2 \xi_\GB (H^2+{\dot H}) \bigg]+8 {\dot H} \left( F_{\GB}+4(H^2+{\dot H}) \xi_{\GB} \right).
\end{align}
Exactly as with the effective gravitational constant, $\Sigma$
approaches a constant at large distance and decays in proportion to
$1/k^2$ at short distance.

\section{Special cases}
The consequences for generic $f(R,\GB)$ theories derived in the last
section do not apply if the model satisfies a condition,
\begin{equation}
{\dot H}^2 \left( F_R \xi_{\GB}-F_{\GB}^2 \right)=0. \label{cond}
\end{equation}
For example, $f(R)$ and $R+f(\GB)$ theories belong to this class.
However, there are infinite number of other theories which satisfy
this condition \cite{DeFelice:2009ak}.  If this condition is
satisfied, $B_3$ vanishes exactly.

Interestingly, (\ref{cond}) is exactly the same condition as the one
for the absence of $k^4$-term in the dispersion relation for the fast
oscillating modes \cite{DeFelice:2010sh}.  If the theory satisfies
(\ref{cond}), the dispersion relation becomes $\omega_1^2 = A k^2$,
where explicit form of $A$ depends on the theory.  For example, for
$f(R)$ theories, $A=1/a^2$.  Therefore, the fast oscillating modes
propagate with a velocity of light and the sub-horizon approximation
we have used can be still applied.  For $R+f(\GB)$ theories, however,
it was found in \cite{DeFelice:2009rw} that modes corresponding to the
fast oscillating ones become highly unstable on small scales in the
radiation/matter dominated era.  In such a case, the unstable modes
cannot be neglected and hence we should include them in the
perturbation analysis.  In this paper, we have assumed that modes
corresponding to the fast oscillating ones in the theory satisfying
(\ref{cond}) are stable and safely decouple from the other modes.

For the special case where (\ref{cond}) holds, the effective gravitational constant approach finite values for large $k$.
Decay of $Q$ that we have observed in the last section does not occur for the special case.
Correspondingly, the gravitational potential (\ref{poten}) around a mass point in this case is given by
\begin{equation}
\Psi ({\vec x},t)= -\frac{G_NM}{r} \frac{A_1}{B_1} \bigg[ 1+\frac{B_1 A_2-A_1 B_2}{A_1 B_2} \exp \left( -\sqrt{\frac{B_1}{B_2}} ar \right) \bigg].
\end{equation}
The second exponential term acts as shifting the gravitational constant by a factor $B_1 A_2/(A_1 B_2)$ 
as we go down to the smaller scales.
This exponential type of gravitational potential was also derived for $f(R)$ theories \cite{Olmo:2005jd,Chiba:2006jp,Navarro:2006mw}.

The power-law decay at short distance is also absent for $\Sigma$.
We see that it asymptotically approaches $C_2/B_2$ at short distances.
In the following, just for the purpose of demonstration, we will give expressions of $Q$ and $\Sigma$
for $f(R)$ and $f(R+\GB/ \Lambda^2)$ theories.

\subsection{Special case I : $f(R)$ theories}
Let us first consider $f(R)$ theories. In this case, we have
\begin{align}
&Q=\frac{1}{F} \frac{1+4 \frac{F_R}{F}\frac{k^2}{a^2} }{1+3 \frac{F_R}{F}\frac{k^2}{a^2}},\\
&\Sigma=\frac{1}{F}.
\end{align}
Therefore, we have $\Sigma=Q$ at large scales and $\Sigma=\frac{3}{4} Q$ at short scales.

\subsection{Special case II : $f(R+\GB/ \Lambda^2)$ theories}
Let us next consider $f(R+\GB/ \Lambda^2)$ theories, where $\Lambda$ is a constant of mass dimension.
We can easily check that this model satisfies (\ref{cond}).
In the limit $\Lambda \to \infty$, the theory reduces to the $f(R)$ theory.

In this model, $\Sigma$ and $\mu$ are given by
\begin{align}
&Q=\frac{\Lambda^2}{F} \frac{\Lambda^4+4 {(\Lambda^2+4H^2+4{\dot H})}^2 \frac{F_R}{F}\frac{k^2}{a^2}}{\Lambda^6 F+ (\Lambda^2+4H^2)W \frac{F_R}{F}\frac{k^2}{a^2}}, \\
&\Sigma=\frac{\Lambda^2}{F} \frac{\Lambda^4 + (\Lambda^2+4H^2+4{\dot H})(3\Lambda^2+12H^2+8{\dot H})\frac{F_R}{F} \frac{k^2}{a^2}}{\Lambda^6 F+ (\Lambda^2+4H^2) W\frac{F_R}{F}\frac{k^2}{a^2}},
\end{align}
where $W$ is defined by
\begin{equation}
W=16H {\dot F}(\Lambda^2+4H^2+4{\dot H})+\Lambda^2 F(3\Lambda^2+12H^2+16{\dot H})-4(\Lambda^2+4H^2){\ddot F}. 
\end{equation}
Therefore, we have $\Sigma=Q$ at large scales and 
\begin{equation}
\Sigma = \frac{3\Lambda^2+12H^2+8 {\dot H}}{4(\Lambda^2+4H^2+4{\dot H})}Q,
\end{equation}
at short scales.

\section{Discussion and Conclusion}
We have shown that in $f(R,\GB)$ theories, the effective gravitational
constant generically decays in proportion to $k^{-2}$ on sub-horizon
scales.  This means that the gravitational force between the two
massive objects is weaker than the one in GR and such a deviation will
be larger on smaller scales.  Then, we naively expect that the
strongest constraint on $f(R,\GB)$ theories will be obtained either
from the solar system constraints or Cavendish type experiments on the
Earth.  However, the direct application of our result to such small
scale systems is not justified since the background spacetime is no
longer FLRW universe.  Instead, the background spacetime we should use
for the solar system, for example, is the spherical symmetric static
metric that must be close to the Schwarzshild metric. 
Just to illustrate this issue more in detail,
we will consider here, as one example, the solar-system constraints on
the dark energy model \cite{DeFelice:2009aj}:
\begin{equation}
f(R,\GB)=R+\lambda \frac{\GB}{\sqrt{\GB_*}} \arctan \left( \frac{\GB}{\GB_*} \right)-\alpha \lambda \sqrt{\GB_*},
\end{equation}
where $\alpha,\lambda$ and $\GB_*$ are positive constants, was studied (many
other models were also analyzed in \cite{DeFelice:2009aj}).
The Gauss-Bonnet correction to GR is highly-suppressed in the solar-system 
by $\epsilon \equiv \sqrt{\GB_*}/\sqrt{\GB_s} \ll 1$ where $\GB_s$ is the Gauss-Bonnet
term at the solar-system.
It was shown that the spherical-symmetric metric:
\begin{equation}
ds^2=-A(r) dt^2+\frac{1}{B(r)} dr^2+r^2 d\Omega^2,
\end{equation}
can be expanded in $\epsilon$ as,
\begin{eqnarray}
A(r)=1-\frac{r_s}{r}+\epsilon c_1 {\left( \frac{r}{r_s} \right)}^p+{\cal O}(\epsilon^2), ~~~~~B(r)=1-\frac{r_s}{r}+\epsilon c_2 {\left( \frac{r}{r_s} \right)}^q+{\cal O}(\epsilon^2), \label{expansion}
\end{eqnarray}
where $c_1,c_2,p$ and $q$ are model-dependent ${\cal O}(1)$ constants.

For this kind of background, it is clear that the effective gravitational constant 
is close to the Newton's constant and we do not see the $k^{-2}$ scaling.  
Therefore, the diminishing of the gravitational constant we have observed occurs only
when the objects whose background spacetime can be treated as FLRW
universe, that is, when the expansion (\ref{expansion}) breaks down \footnote{
Precisely speaking, this model gives $\Xi=0$. However, the expansion (\ref{expansion}) is
quite general and we expect the similar conclusion can be reached for the general $f(R,\GB)$ theories as well,
such as models of the kind $f(R,\GB)=a/(R^p+b \GB^q)$. }. 
For some models, the distance $r$ at which the expansion breaks down can be
much smaller than the horizon scale (But it must be still larger than the solar-system
size for such a model to be viable.). 
In such a case, there may be a region of distance scales where the weakening of gravity
may be significant and its signature may be
imprinted in the patterns of the cosmic structures such as group of galaxies as less structures
than GR, which can be tested by observations.  In principle, we can
construct $f(R,\GB)$ theories that possess those properties by
requiring that $f(R,\GB)$ becomes very close to GR for large $R$ and
deviates from GR when $R$ and $\GB$ are ${\cal O}(H_0^2)$ and ${\cal O}(H_0^4)$, where $H_0$ is the Hubble constant.

We also found that if the theory satisfies ${\dot H}^2 \Xi=0$, 
then the weakening of gravity does not
occur.  Interestingly, exactly the same combination appears in the
dispersion relation for the fast oscillating modes.  We conclude that
the short scale behavior of the perturbations and the gravity are
qualitatively dependent on whether ${\dot H}^2 \Xi$
vanishes or not.  If, at the scales where perturbation theory can be
applied to study the data, this quantity does give some non negligible
contribution, then we expect the matter spectrum, at the same scales,
to be in general suppressed.

\section*{Acknowledgements}
We would like to thank Jean-Marc G\'erard and Shinji Tsujikawa for
  helpful discussions. The work of A.\,D.\,F.\ was supported by the
  Grant-in-Aid for Scientific Research Fund of the JSPS No.\ 09314.
  T.\,S.\ is supported by a Grant-in-Aid for JSPS Fellows No.\
  1008477.  This work is supported by the Belgian Federal Office for
  Scientific, Technical and Cultural Affairs through the
  Interuniversity Attraction Pole P6/11.

%

\end{document}